\documentclass[sigconf]{acmart}

\usepackage{booktabs} 

\usepackage{graphicx}

\usepackage{amssymb}
\usepackage{amsmath}

\usepackage{dsfont}

\usepackage{caption}
\usepackage{subcaption}

\usepackage{tabulary}

\usepackage{threeparttable}

\usepackage[linesnumbered]{algorithm2e}

\begin{document}

\title{On Identifying Anomalies in Tor Usage with\\ Applications in Detecting Internet Censorship}

\author{Joss Wright}
\affiliation{%
	\institution{University of Oxford}
	\city{Oxford}
	\country{United Kingdom}
}
\email{joss.wright@oii.ox.ac.uk}

\author{Alexander Darer}
\affiliation{%
	\institution{University of Oxford}
	\city{Oxford}
	\country{United Kingdom}
}
\email{alexander.darer@linacre.ox.ac.uk}

\author{Oliver Farnan}
\affiliation{%
	\institution{University of Oxford}
	\city{Oxford}
	\country{United Kingdom}
}
\email{oliver.farnan@balliol.ox.ac.uk}

\begin{abstract}

	We develop a means to detect ongoing per-country anomalies in the
	daily usage metrics of the Tor anonymous communication network, and
	demonstrate the applicability of this technique to identifying likely
	periods of internet censorship and related events. The presented
	approach identifies contiguous anomalous periods, rather than daily
	spikes or drops, and allows anomalies to be ranked according to
	deviation from expected behaviour.

	The developed method is implemented as a running tool, with outputs
	published daily by mailing list. This list highlights per-country
	anomalous Tor usage, and produces a daily ranking of countries according
	to the level of detected anomalous behaviour. This list has been active
	since August 2016, and is in use by a number of individuals, academics,
	and NGOs as an early warning system for potential censorship events.

	We focus on Tor, however the presented approach is more generally
	applicable to usage data of other services, both individually and
	in combination. We demonstrate that combining multiple data sources
	allows more specific identification of likely Tor blocking events. We
	demonstrate the our approach in comparison to existing anomaly detection
	tools, and against both known historical internet censorship events and
	synthetic datasets. Finally, we detail a number of significant recent
	anomalous events and behaviours identified by our tool.

\end{abstract}

\ccsdesc[500]{Networks~Network measurement}
\ccsdesc[500]{Social and professional topics~Technology and censorship}
\ccsdesc[300]{Security and privacy~Pseudonymity, anonymity and untraceability}

\keywords{information controls, censorship, filtering, anomaly detection}

\setcopyright{none}
\settopmatter{printacmref=false} 
\renewcommand\footnotetextcopyrightpermission[1]{} 
\pagestyle{plain} 

\maketitle


\section{Introduction}
\label{introduction}

Nation states, and others, increasingly employ internet filtering as a means of
controlling access to information, and as a tool to limit social and political
organisation.  Given the central role that the internet plays in communications
for a large proportion of the global population, understanding the application
and development of filtering technologies, and the effects of these methods on
individuals and society, is of great importance.  Whilst analyses of known
filtering regimes allow us to identify tools, techniques, and limitations of
filtering approaches, we consider that discovering internet filtering behaviour
in less-studied regions is of great importance.

Much existing research into internet filtering has focused either on observing
practices of states already known engage in filtering, or in the development of
censorship circumvention tools.  Whilst multilateral studies of censorship have
been conducted, most notably the seminal work of Deibert et al.
\cite{deibert2009denied}, these approaches have typically amalgamated manual
country-specific investigations. In the case of Deibert et al.,
countries were hand-ranked according to a number of broad criteria for internet
freedom, based on network measurements as well as media reporting and expert
interviews.

The work presented here provides a means to alert researchers and activists to
developing events that may otherwise have been missed by focusing on patterns
of circumvention tool usage around the world. As an initial step our tool
currently reports new anomalies and a current ranking of most anomalous
countries to a mailing list on a daily basis. The \texttt{<infolabe-anomalies>}
mailing list has been running publicly since August 2016, has subscribers from
academia and civil society organisations, and has provided the first known
detection of a number of significant ongoing Tor-related blocking events that
we detail in \S\ref{section:examples}.



\subsection{Contributions}

This work presents a theoretical contribution to network anomaly detection, a
practical contribution in the form of an implemented tool for detecting
anomalous events in Tor usage data, a resource in the form of a public dataset
of detected anomalies in historical Tor traffic, and a practical analysis
demonstrating the detection of real-world events: we identify known, previously
unreported, and newly-detected filtering-related events.

We make the following practical contributions:
\begin{itemize}

	\item An open tool to detect and highlight anomalies in per-country usage of the Tor network;

	\item a continually-updated daily ranking of the most anomalous countries in terms of their usage of Tor.

\end{itemize}

These are built on our key methodological contribution:
\begin{itemize}
	\item An approach for detecting and quantifying anomalous periods of per-country Tor usage incorporating multiple usage measurements.
\end{itemize}

We validate the effectiveness of our approach in detecting both a range of
artificial anomalies, and known reported filtering events against the Tor
network. We also demonstrate our approach's improved detection accuracy
compared to the existing Tor metrics anomaly detector, as well showing its
additional capabilities in terms of identifying anomalous periods and ranking
anomalies by strength.


\subsection{Problem and Approach}
\label{approach}

When an entity, such as a state or ISP, chooses to filter or block certain
types of information, the resulting patterns of traffic reflect the
intervention in the form of statistical anomalies. In a global system, in which
many entities may be interfering with traffic or publicising their attempts to
do so, it is desirable to identify \textit{localised} anomalies and to gain an
understanding of their nature.


To detect anomlies, we model each country's Tor usage \textit{relative} to the behaviour of other
countries, not as an individual time series. A given country's usage pattern is
judged as anomalous if it deviates from its previous behaviour \textit{relative
to other countries}. 




The usage patterns of a tool such as Tor, explicitly developed and publicised
as a means for bypassing network censorship, are affected by a range of factors
such as filtering, social and political unrest, unrelated network outages, and
media reporting~\cite{179502}. The work presented here therefore identifies
\textit{statistical anomalies} in Tor usage metrics, but we highlight that such
anomalies serve as an indicator, not a proof, of censorship or interference.



In later sections we make use of both standard Tor traffic and
blocking-resistant \textit{bridge node} traffic to identify direct blocking of
Tor. Combining anomalies across metrics allows identification of declines
in normal usage combined with rises in blocking-resistant bridge usage. This
corresponds to users being unable to access Tor normally, and so switching to
blocking resistant approaches. As we demonstrate in \S\ref{section:examples},
this provides a targeted identification of filtering-related anomalies.



We extend a line of research initially proposed by Jackson and Mudholkar
\cite{jackson1979cpr} for application in industrial process control, and later
employed by Lakhina et al.~\cite{LakhinaCrovellaDiot:sigcomm2004} to detect
network-wide traffic anomalies from per-link data in high-performance networks.
Our approach differs from that of \cite{LakhinaCrovellaDiot:sigcomm2004} in a
number of ways. Firstly, we do not assume that the underlying set of time
series are stationary, but instead allow for series to evolve over time.
Secondly, we account for \textit{seasonality} in time series. Most importantly,
however, we identify \textit{per-country} anomalies rather than global.
Finally, we dynamically adapt our anomaly thresholds for each series to account
for long-term evolution of the data.

We directly apply our tool to analysis of Tor usage anomalies, and report on
its demonstrated utility for detecting anomalies of practical concern to
activists and NGOs working to support censorship circumvention and freedom of
expression. A number of such actors subscribe to our public mailing list, and
have used our detection results to identify newly-emerging filtering
behaviours.

\section{Existing Work}

Internet filtering has received attention from various fields. Technical
research has focused on mechanisms of censorship and the development of
circumvention approaches. The social sciences have investigated motivations of
censors, and their legal, economic, and societal effects.


\subsection{Technical Analysis}

Arguably the most well-known national-level filtering system is that of China,
commonly known as the Great Firewall. One of the earliest significant
studies of this system was presented by Clayton et
al.\cite{clayton2006ignoring}, who isolated one mechanism by which
connections were interrupted if particular keywords were identified in traffic.
The mechanism discovered by Clayton et al. resulted in TCP RST packets being
sent from an intermediary router to both source and destination of a connection
if a filtering criterion was met. The authors further demonstrated that
if the two endpoints of the connection ignored the TCP RST, the connection
could successfully continue.

In more recent work, it has become apparent that the Chinese approach to
filtering is both complex and evolving. In two recent papers, a group of
anonymous researchers have explored manipulation, or poisoning, of DNS records
that pass through China
\cite{anonymous2012collateral,anonymous2014comprehensive}. This work has
identified DNS manipulation as one of the most prevalent forms of filtering in
China. Similarly, Wright \cite{wright2012regional} demonstrated that DNS
censorship had different effects between different regions within China, with
significant variation in the nature of the DNS poisoning seen across the
country. Similarly, Farnan et al. \cite{farnan2016poisoning} showed that the
approach taken to DNS poisoning in China resulted in pollution of both network
requests and DNS servers themselves.

Crandall et al.\cite{crandall07concept} make use of \textit{latent semantic
analysis} to derive, from known terms blocked in HTTP traffic going into China,
semantically related keywords that might also be blocked. These derived
keywords can then be verified by the simple process of attempting to make HTTP
connections into China containing the suspect words.  This approach aims to
produce a continually-updated list of blocked terms that could be used to
maintain an understanding of those terms most offensive to the filtering
authorities. Similarly, Darer et
al.~\cite{darer2017keywords,2018arXiv180403056D} have used keyword- and
crawling-based approaches to discover previously unindentified blocked domains. 

\subsection{Global Studies}

Perhaps the most comprehensive study to date of global filtering practices is
given by Deibert et al.~\cite{deibert2009denied}. In this work the authors
carried out a range of remote and in-country analyses over a number of years,
incorporating both technical measurements and interviews with local experts.
The resulting research presented a series of snapshots of individual countries,
with both an overview of the social, political, and technical landscape, and
censorship practices rated on a simple scale in various categories of content:
political, social, conflict and security, and internet tools.

%
%
Some forms of filtering act not at the network layer, but on application level or
social filtering. King et al.~\cite{king2013expression} studied manual
censorship practices in Chinese long-form blogging, and demonstrated that the
Chinese censorship authorities were chiefly concerned with preventing calls to
\textit{collective action} whilst allowing significant levels of government
criticism.

\subsection{Anomaly Detection}

The Tor project maintain a censorship flagging tool, as described by
Danezis\cite{Danezis2011a}. This tool uses a particle-filtering approach to
model the ratio of daily connections for each country in a seven-day time
period. If a country's ratio of current to past users increases or decreases
significantly more than the average of the fifty largest Tor-using countries,
then an anomaly is flagged. These reported anomalies are available at the Tor
Project's metrics portal~\cite{tor-metrics}. We evaluate our approach's
accuracy against that of Danezis in \S\ref{section:validation}.

A related approach was used by Lakhina et
al.~\cite{LakhinaCrovellaDiot:sigcomm2004} to identify \textit{network-wide}
anomalies in high-speed networks. This work assumed that long-term network
usage was stable, and made use of data gathered from a restricted set of
link-level observation points to detect network-wide anomalies. Our approach
relaxes both of these assumptions, neither of which hold for the Tor metrics
data. These extensions are discussed in greater detail in
\S\ref{rolling-analysis}.

Several other works have extended or expanded aspects of
\cite{LakhinaCrovellaDiot:sigcomm2004}, notably
\cite{Soule05combiningfiltering}, \cite{Zhang05networkanomography}, and
\cite{Huang07communication-efficientonline}. These largely focus, however, on
using a small number of network observation points to infer network-wide
anomalies, and as such typically begin from relatively low-dimensional data.
Our approach inverts this concept by detecting per-observation anomalies across
a dataset with several hundred dimensions, representing individual countries'
usage, in order to highlight states displaying anomalous behaviour.


\section{Concepts}

In this section we discuss the fundamental techniques underlying our approach,
and discuss their application to the dataset we use in the rest of this work.


\subsection{Tor}

Tor~\cite{dingledine04tor} is an approach to anonymous web-browsing that
offers realistic compromises between latency, usability, and the strength of
the anonymity properties that it provides. The most visible end-user aspect of
Tor is the Tor Browser Bundle, which provides a web-browser that both uses the
Tor network for transport, and is tailored to reduce identifiability of end
users.

Managed by the Tor Project, Tor has developed into a global network of
volunteer-run relays that forward traffic on behalf of other users. The network
makes use of an \textit{onion routing} approach that build encrypted circuits
between relays, preventing most realistic adversaries from linking Tor users to
particular streams of traffic exiting the network.

The most sigificant aspect of the Tor network for the present work is that, by
its nature, users' traffic is relayed via third parties. As such, and in
addition to its anonymity properties, Tor provides a means to bypass many forms
of internet filtering. Censorship circumvention is a core aspect of the Tor
Project's goals, and significant ongoing research
work\cite{ccs2012-skypemorph,ccs2012-stegotorus,Shirazi:2015:TMR:2808138.2808152}
is aimed at ensuring that Tor is resilient against attacks and continues to
offer means to evade national-level filters.

While the extent and popularity of Tor's use in regions that experience
significant levels of filtering, such as China, is open to debate
\cite{yu2013collateral}, Tor is known to have been blocked actively by a number
of states, including China and Iran, that object to its use to bypass local
internet restrictions and to act anonymously. Significantly, Tor is also
arguably the highest-profile censorship circumvention tool at the international
level and has received significant media coverage, making it one of the tools
of choice for internet activists.

\subsubsection{Tor Metrics Data}

Tor's role as a high-profile censorship circumvention network make it a useful
indicator of global filtering practices. To support analysis of the tool, the
Tor project provide estimated daily per-country usage statistics, gathered by
counting the number of client requests to central \textit{directory
authorities} on a daily basis.

%
%
It is assumed that each client, on average, will make ten requests per day, and
as such the aggregate user statistics are divided by ten to provide a final
estimate of usage. This data is averaged across each 24-hour period to provide
the average number of concurrently connected Tor clients for that
day\cite{tormetrics-qa}. Whilst the number of distinct clients per day cannot
be estimated with any accuracy, the methodology of the Tor metrics portal
provides a sufficiently stable estimate.

From these estimates we obtain a set of 251 time series representing individual
countries according to the GeoIP database used by Tor. These time series
comprise daily observations ranging from the beginning of September 2011 to the
time of writing. From these, we remove those countries whose Tor usage never
rises above 100 users to discount countries whose variance is too high to allow
meaningful anomaly detection. 

In later sections, we combine normal usage trends in Tor with censorship-resistant \textit{bridge node} usage to identify correlated anomalies. This is discussed in further detail in \S\ref{section:combined}.

%



\subsection{Principal Component Analysis}
\label{pca}

Principal component analysis was developed by Pearson\cite{pearson1901lines} as
a means to produce tractable low-dimensional approximations of high-dimensional
datasets. The original set of variables, which may display correlations, are
transformed to a set of linearly uncorrelated variables know as
\textit{principal components}.



When data displays a high degree of correlation between variables then a small
number of the most significant principal components may be sufficient to
describe the original data to a high degree of accuracy. In many practical
scenarios, high dimensional data can be described using only two or three of
the most significant principal components. See \cite{jolliffe2002principal} for
a detailed treatment of principal component analysis and the various choices
and compromises to be made when applying the technique.

%

The practical result of this is that our results are not influenced by
countries with large usage numbers; the principal component analysis considers
variance, not magnitude, in calculating the contribution of each country to the
model.

\section{Approach}

The basic operation of our approach are described here, and are given as
pseudocode in Algorithm~\ref{alg:pseudocode}.

\begin{algorithm}
	\caption{Basic anomaly tagging algorithm. (Anomaly magnitudes omitted for brevity.)} \label{alg:pseudocode}
	\DontPrintSemicolon
	\SetKwInOut{Input}{input}
	\SetKwInOut{Output}{output}
	\SetKwProg{PCATagAnomaly}{PCATagAnomaly}{}{}

	\PCATagAnomaly{}{
		\Input{\textit{usage} $\gets$ Set of per-country time series}
		\Output{\textit{anomalies} $\gets$ Set of per-country anomaly time series}
		(Clean data; remove seasonality)\;
		$medians \gets$ \{median residual errors for each country\}\;
		$mads \gets$ \{median absolute deviations (MADs) of residual errors for each country\}\;

		\ForEach{day in usage}{
			$pc \gets$ calculate principal components over all countries' usage[(day-179):day]\;
			\ForEach{country}{
				$recons \gets$ reconstruct $day$ value for $country$ using $pc[1:12]$\;
				$obsv \gets$ observed value for final day for $country$\;
				$err \gets$ abs( $obsv$ - $recons$ ).\;
				$medians_{country} \gets$ update median using $err$\;
				$mads_{country} \gets$ update MADs using $medians_{country}$ and $err$\;
				\If{abs($err$) $>$ abs( $mads_{country} \times 2.5$ ) }{
					$anomalies_{country \times day} \gets$ 1\;
				}
				\Else{
					$anomalies_{country \times day} \gets$ 0\;
				}
			}
		}
	}
\end{algorithm}

\subsection{Overview}

Starting from Tor's per-country usage data, we initially remove all countries
whose usage never rises above 100 users, to avoid the unacceptably high
variance in such data. We then apply the STL algorithm to identify and remove
any seasonality -- in our case weekly trends -- in individual countries. 

For each 180-day period in the dataset we apply a principal component analysis
over the usage time series for all countries, resulting in a set of components
for that time window. Taking the true observed usage for each country for the
final day of each window, we calculate the \textit{approximated} value from the
first 12 principal components. This provides the expected value for each
country based on previous behaviour\footnote{Using the full set of principal
	components at this stage would result in a perfect reconstruction of the
original observed values.}.

For each country we calculate the difference between the true value and the
reconstructed value, providing a \textit{residual error} that was not captured
by the restricted set of principal components.

We maintain a rolling calculation of both the median observed residual error
and the median absolute deviation of the errors for each country. We mark a day
as anomalous if the observed residual error falls outside of 2.5 median
absolute deviations from the median.

We now detail the individual steps listed above, and justify our choices of
parameters.

\subsubsection{Removal of Seasonality}

Per-country Tor usage data, as with much network usage data, exhibits
significant \textit{seasonality}, typically on a weekly basis, reflecting
changes between usage on weekdays and at weekends. This continual cyclical
change in usage can reduce the accuracy of principal component analysis due to
varying levels of seasonality exhibited by different countries.

We employ the \textit{Seasonal and Trend Decomposition using Loess} (STL)
method of Cleveland et al.~\cite{ClevelandEtAl:1990} to remove the seasonal
component of each series, leaving the trend component and the residual noise as
inputs to our anomaly detector. In later sections, however, we show the
original data with seasonality restored.

\subsubsection{Rolling Analysis}
\label{rolling-analysis}

Principal component analysis does not account for ordering in observations, and
as such cannot account for evolution of a dataset according to trends or
seasonablity. To account for developing patterns, therefore, we perform a
rolling principal component analysis over smaller time windows within the
series. For the purposes of our experiments, we make use of a 180-day window as
a balance between sufficient data for useful principal component analysis,
given the number of dimension in the data, against the evolution of the daily
Tor metrics. See Ringberg et al.~\cite{ringberg2007sensitivity} for a
discussion of the sensitivity of PCA to such factors.

\subsubsection{Selection of Components}



For PCA, the full set of principal components allows reconstruction of the full
data set. As fewer components are selected, less variance in the original
dataset is captured. A common approach to selecting an appropriate number of
components for modelling is to make use of \textit{Kaiser's
criterion}~\cite{kaiser1960factor} to select only those principal components
with eigenvalue greater than 1, representing those components that provide more
information than a single average component. Based on this heuristic, our
experimental results suggest twelve principal components as broadly optimal
across the dataset. 

%



With appropriately calculated principal components, we can reconstruct an approximate value for each day's Tor usage based on previous behaviour. We highlight that at no point do we \textit{predict} forecasted values for usage in future days. In each case, we reconstruct a day's usage based on principal components in order to compare against the true observed  value, and thus to calculate deviance from prior behaviour relative to other countries.

\subsection{Calculation of Residuals}

After reconstructing data from principal components, the result is a set of
\textit{residuals} that express variances in the observed data not captured by
the current principal component model. A sufficiently large-scale residual
represents behaviour that deviates significantly from previous patterns, and is
thus of interest.

\subsection{Identifying Anomalies through Residuals}

The residual errors calculated during the reconstruction accounts for variance
in the dataset that is not expressed by the chosen principle components in the
approximate model. 


\begin{itemize}
	\item Positive residuals represent drops in expected Tor usage for a
		country.

	\item Negative residuals represent increases in expected Tor usage for a
country.

	\item Magnitude of residuals expresses how much a country varies
from its previous behaviour relative to other countries.

\end{itemize}


A key advantage of identifying anomalies from residual errors rather than raw
usage numbers is that it incorporates the expected \textit{trend} of the data.
This identifies anomalous periods even when no visible shift in usage is seen:
a flat usage trend where the expectation is a rise or fall is correctly
identified as anomalous by our approach. This capacity to identify anomalies in
apparently typical usage is an important and unusual aspect of our technique,
taking advantage of the relative patterns of usage between countries. 

A second advantage of this approach is that each day can be judged as anomalous
or not based on a model of behaviour relative to other countries. As such, in
contrast to many other anomaly detection approaches, we identify
\textit{periods} of anomalous behaviour in which a country may be experiencing
ongoing elevated or reduced usage. Other approaches typically flag an
individual day as a significant spike or drop, but cannot identify ongoing
periods as anomalous. This capability greatly aids our ability to study
time-bounded changes in Tor usage.


%

\subsection{Combining Features to Identify Targeted Filtering}
\label{section:combined}

It is fundamental to the broader goals of this work that usage anomalies in
appropriately selected traffic, and in particular from circumvention tools, can
be indicative of the imposition or relaxation of filtering. At the same time, it
is clear that other types of event, both technical and sociopolitical, can lead
to shifting patterns of usage in these tools. 

We aim to identify two forms of event: firstly, direct blocking of the Tor
network; secondly, changing characteristics of Tor usage in response to
exogenous factors. The censorship of a major international website, such as
YouTube, has the potential to drive a noticeable number of users to Tor, and as
such Tor becomes a useful \textit{proxy variable}~\cite{upton2002oxford} for a
broader class of filtering behaviour. We discuss this in relation to specific
events in \S\ref{section:examples}.

For the first of these classes of event, we detect likely candidates by
carrying out anomaly detection on multiple metrics and combining outputs to
highlight periods in which anomalies were detected in more than one series. The
most useful of these for our purposes is to combine negative trends in standard
Tor usage with positive trends in blocking-resistant bridge node usage,
reflecting users unable to access Tor normally switching to the tool's
blocking-resistant mode.



As such we can identify days in which both standard and blocking-resistant time
series were anomalous. Even without refinements, such as allowing time lags
between anomalies in each series, this approach already highlight a number of
significant cases, which are illustrated in \S\ref{section:examples}.

\subsection{Expected Error and Anomalous Threshold}

A key element in the approach presented in this work is to determine an
appropriate threshold for events to be considered anomalous. The size of this
threshold value is inherently linked to the expected error in the technique. We
here discuss and justify our approach to calculating this threshold, making use
of \textit{robust statistics}\cite{huber2004robust} to minimise false detection
rates.



\label{dynamic-threshold}

A na\"{i}ve anomalous threshold can be defined as a proportion of the usage for
that day. If the reconstructed value deviates by more than some percentage of
the observed value, an anomaly is detected.

This approach is problematic. Critically, different countries may be modelled
more or less accurately than others. As such, countries that are typically
modelled poorly would produce a high proportion of anomalous periods.

As such, we calculate an ongoing threshold based on the characteristics of each
country. By tracking the expected residual value for each country an expected
anomalous threshold can be determined based on typical observed errors.



The standard approach of basing this threshold on the mean and standard
deviations are, however, not \textit{robust} against outliers in the dataset
due to their assumption that errors are Gaussian. We therefore calculate
thresholds based on the \textit{median absolute deviation about the median}
(MAD) to define the expected error in normal usage~\cite{leys2013deviation}.

The median is robust against outliers in the dataset; a small number of extreme
events do not significantly alter its value. Similarly, by taking the median of
the absolute deviations about the median as a measure of the statistical
dispersion in the dataset, we avoid anomalies from overly affecting the
remaining data points.

As a default, we consider events as anomalous if they fall outside of 2.5
median absolute deviations\footnote{Corresponding to roughly one expected false
positive every 80 days. See \S{\ref{false-positives}} for an experimental
analysis of false positives in our approach.} from the rolling median value.
See \cite{leys2013deviation} for a discussion of the robustness of the median
and MAD against outliers, and a justification of a 2.5 median absolute
deviation threshold.

\subsection{Ranking of Countries}

\begin{figure}
\centering
\includegraphics[width=0.48\textwidth]{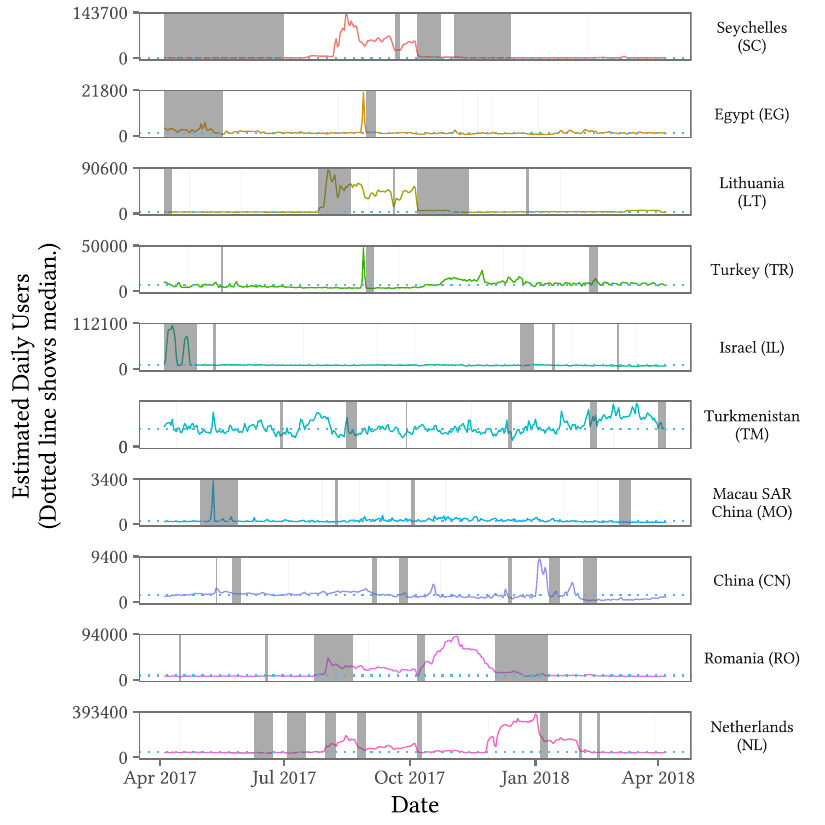}
\caption{Ten most anomalous countries according to median absolute deviation of residuals over the previous year. Grey areas highlight detected anomalous periods.} 
\label{fig:top-10}
\end{figure}

The size of the residual error from the principal componenet analysis provides
a convenient metric by which to rank countries according to the level of
anomalous behaviour that they exhibit in a given time period. We make use of
the size of the median absolute deviation about the median to rank countries,
as shown in Figure~\ref{fig:top-10}.

We now proceed to discuss the application of our technique, and the validation
of the approach.

In \S\ref{section:validation} we evaluate our approach against synthetically
injected anomalies in the data to analyse the effectiveness of our detection
methods as the magnitude and severity of the anomalies vary. We also compare
our detection mechanism against the small number of verified reported blocking
events against the Tor network. 

Finally, in \S\ref{section:examples} we conduct a series of
analyses of the Tor metrics data to identify anomalous countries and specific
periods of anomalous behaviour.

\section{Ethics}
\label{section:ethics}

Conducting research into network filtering presents a number of ethical
issues~\cite{wright2011ethics}. The most significant of these is that approaches
to investigating network filtering may require direct access to filtered
networks.  In practice this often involves the participation of in-country
experts to conduct local network tests.

Due to the uncertain legal, or quasi-legal, status of violating or
investigating state-level network filters, it is generally impossible to
quantify the risks to research participants in carrying out network tests. The
classic models of informed consent used in many other fields of research can be
difficult to apply for a number of reasons, the most important of which is the
lack of meaningful informed consent afforded by automated testing on behalf of
users, and the legal uncertainty surrounding attempted access to filtered
resources on a test subject's network connection.

We therefore assert that, where possible, research into network filtering
should make use of passive measurements and existing available data sources.
The work in this paper is a deliberate attempt to maximise the effectiveness of
such a passive approach.

\section{Validation}
\label{section:validation}

In this section, we judge the efficacy of our method in terms of its ability to
detect anomalies, in a variety of circumstances, as well as its false
classification rate.


\label{false-negatives}
\label{false-positives}

A significant dificulty in validating unsupervised machine learning systems
is that it is largely impossible to obtain comprehensive ground truth for
internet filtering events, nor are there publicly-available exhaustive lists of
filtering events. Indeed, the work here was motivated partially in an attempt
to allow a more exhaustive tracking of such events.  Filtering is, by and
large, an opaque process that is rarely announced. Even when states do choose
to filter connections openly, the details of that filtering are not typically
made public.



As observed in \cite{goix2016unsupervised}, this is an inherent problem in
\textit{unsupervised} anomaly detection algorithms. In the the following
sections we address this lack by injecting artificial anomalies into a synthetic
dataset and comparing this to the Tor Project's existing anomaly detection
approach, as well as evaluating our method against an existing list of known
filtering events.

In the following, we examine both false positive and false negative rates in
evaluating detection rates of anomalous behaviour. A false positive in this
context is a period in which there is no genuinely anomalous activity, but
anomalous activity is reported. A false negative is a period in which there is
anomalous activity but is is not detected.


\subsection{Evaluation in Synthetic Data}
\label{detection-of-injected-anomalies}

To test our approach, and to create a fair comparison against the existing
deployed tool from the Tor Project, we inject artificial anomalies into
synthetic data generated according to underlying features of real-world Tor
usage. 

An alternative test for false negatives is to compare the results from our
approach with an external list of known censorship events. This allows us to
test whether periods exist in which we did not detect anomalous behaviour
during a period where external sources believe an event occurred. We take this
approach in \S\ref{known-event-detection}.

\subsection{Generating Synthetic Data}

To evaluate our approach against an approximation of real-world data, we use
the underlying features of genuine observed Tor data to generate a synthetic
set of time series.

To do so, we select a year-long period of Tor data in which no major global
events can be observed. This was to avoid an unfair basis for comparison
between our approach and that of the Danezis. As such, we selected the year
running from the 1\textsuperscript{st} January 2014 to the 31\textsuperscript{st} December 2014.

To remove, as far as possible, genuine anomalies from this dataset we first
decompose the series into trend, seasonal, and residual components through use
of the STL algorithm~\cite{ClevelandEtAl:1990}. This allowed us to preserve
seasonal properties of the data separately from the underlying trend. We
emphasise that, whilst STL is also used in our anomaly detection approach, the
application of it here preserves, rather than removes, the underlying features
of the data and thus is not unfairly biasing the synthetic dataset towards our
approach.

The underlying trend data is then smoothed using a 28-day rolling median
average. Due to the robust nature of the median against small outliers, this
approach preserves broad-scale trends in the data whilst removing, as far
as possible, small-scale deviations. Without an objective labelled set of
anomalies we cannot guarantee that no anomalies were preserved in the final
dataset, but a visual inspection did not reveal any significant causes for
concern. 

We then calculate, for each country, the mean and the standard deviation of the
residual errors after the trend and seasonal components have been removed. This
gives a base set of parameters from which to generate random noise to be added
to each series.

To create the final synthetic dataset, we recombine the smoothed underlying
trend data with the seasonal component and add randomised noise. As it is
impossible to characterise the ``true'' noise process without having labelled
anomalies we conservatively add Gaussian noise drawn according to the observed
mean and standard deviation. This provides a ``clean'' dataset without
anomalies, based on real-world patterns of behaviour.

\subsection{Injecting Anomalies}

As with the underlying data, we generate anomalies based on properties observed
in real-world data. The strength of the injected anomalies is based on the
average daily users for each country, and magnified upwards or downwards
gradually to create the anomaly. 

To create an anomaly, the number of users in each set was increased or
decreased by 0--100\%. Anomalies are added to the data gradually, ranging over
periods from one to four weeks. These parameters were selected based on
observation of known anomalies and visual inspection of the original dataset.

In total, for the year of synthetic data, we injected a total of 250 anomalies
across all countries, randomly drawn from the space of possible parameters.

This synthetic, labelled dataset provides the basis both for objective
evaluation of the effectiveness of our technique, and as an unbiased means of
comparison between our approach and that of \cite{Danezis2011a}. We now
evaluate the effectiveness of these two approaches.




\subsection{Comparison of Tools} 

An evaluation of false positive and false negative rates in detecting anomalous
periods allows both an objective judgement on the effectiveness of our
approach, and a comparison against the existing tool used by the Tor
Project~\cite{Danezis2011a}. To carry out this comparison, we formatted the
clean synthetic dataset appropriately for each tool and compared the detected
anomaly series from each to the injected set of anomalies.

One problematic element of such a comparison is in the nature of event
reporting from each tool. As mentioned, our approach reports day-by-day
anomalies based on principal component modelling. By comparison,
\cite{Danezis2011a} bases its detection on significant spikes and dips on a
day-by-day basis. As such, it is far less likely that Tor Project's existing
tool will report anomalous periods, but will instead detect only the points at
at which an anomaly starts and ends. This should hypothetically result in a
much higher detection accuracy rate for our tool on a day-by-day comparison: an
anomaly that lasts for ten days will typically only produce two anomalously
flagged days in the Tor Project's detection scheme, whereas it may result in
ten days for our tool as each day in the anomalous period may be identified. By
contrast, however, our tool's approach leaves us open to a potentially higher
false negative rate when a period is falsely judged to be anomalous.

We highlight again, however, that this period-based rather than event-based
approach is one of the key strengths of our improved approach -- we report
entire periods as anomalous rather than simply identifying point anomalies.

As such, to compare, we perform a simple analysis: the output of each tool is
evaluated according to the ground truth in the labelled synthetic dataset. Days
correctly identified as anomalous contribute to the \textit{true positive}
rate, whilst days marked as anomalous that are not in the synthetic data
contribute to the \textit{false positive} rate. Similarly, if a day is
anomalous in the synthetic data and missed by our tool, it contributes to the
\textit{false negative} rate, whilst days correctly identified as not anomalous
contribute to the \textit{true negative} rate. These values are reported in
Table~\ref{table:comparison-results}.

\begin{table}
	\begin{center}
		\begin{threeparttable}
			\begin{tabulary}{\textwidth}{CCC}
				\hline
								    	&	Tor Metrics 		&		Principal Component \\ 
				\hline
				True Positives 	& 	8.57\%  	& 		20.08\%	\\
				True Negatives 	& 	92.75\% 	& 		94.25\%	\\
				False Positives 	& 	7.25\%  	& 		5.75\%	\\
				False Negatives 	& 	91.43\% 	& 		79.92\%	\\
				\hline 
				Total Days Flagged &	2962		& 		2820 		\\
				\hline
				Minimal Detection Total\tnote{1} & 88 & 139 \\
				\hline
			\end{tabulary}
			\begin{tablenotes}
			\item Total anomalous days across entire set was 4214.
			\item[1] Anomalies during which at least one day was identified.
			\end{tablenotes}
		\end{threeparttable}
	\end{center}
	\caption{Comparison between Tor Metrics and Principal Component approach on synthetic data.}
	\label{table:comparison-results}
\end{table}

Our principal component-based approach significantly outperforms the currently
deployed Tor Metrics detector both in marking genuine anomalies and in avoiding
marking non-anomalous days incorrectly. 

The overall detection rate of our approach is over twice that of the
alternative, at 20\% of all genuinely anomalous days being identified. This
figure is somewhat misleadingly low, however, as this includes many
correctly-identified anomalous \textit{periods} for which, however, some
individual days were not themselves considered anomalous. 

These results suggest that in realistic data generated from observed real-world
trends, the proposed principal component analysis-based approach significantly
outperforms the existing deployed tool.

\subsubsection{Ranking}

We have attempted, as far as possible, to undertake a fair comparison of the
quantitatively comparable elements of these two approaches, despite significant
differences in their output. In addition, however, our approach offers a number
of advantages for analysis. The most significant of these is the ability to
rank countries according to the strength of the anomalies they have
demonstrated over time in terms of deviation from expected behaviour. The
\texttt{infolabe-anomalies} mailing list reports daily the top-ranking
anomalous countries for the previous day, week, and month in addition to a list
of all countries anomalous for that day.

It is worth highlighting that whilst realtime detection is of great interest to
the commmunity, the ability to study historical anomalies in the Tor metrics
dataset is also of significant value.

\subsection{Detection of Known Events}
\label{known-event-detection}

Having calculated anomalous statistics over a synthetic data set, we now aim to
validate our approach by comparing anomalies detected in real data against
countries and periods in which internet restrictions are known to have been
applied, or in which significant events were occurring that may have influenced
usage of circumvention tools.


For this purpose we use \cite{tor-timeline}, a list of reported and verified
filtering events against the Tor network dating from 2008 to 2015. This list
includes a brief description of each reported event, the dates when the event
was first reported, and how the blocking was resolved. 

The list of events used in this evaluation\cite{tor-timeline} was compiled
through bug reports, talks, examination of blog postings, and the use of
machine learning on blog postings to identify reports of censorship
automatically. As such, the exact timing of the events is somewhat fuzzy; a
blocking event against Tor could have occurred some time before bug reports and
blog postings were filed.

In addition, \cite{tor-timeline} is unfortunately brief, reflecting a
significant lack of data available concerning this topic. As discussed, a
motivation for this work is to provide a baseline of reliable indicators to
allow for potentially censorship-related anomalies to be identified and
investigated more thoroughly.

The Tor Project's metrics data does not cover the full time range of the events
listed in \cite{tor-timeline}. For those events that do fall within the
available data, we analyse here whether these would be detected by our
approach.

As shown in Table~\ref{table:reported-events}, only eight reported events coincide
with the available published metrics data. Of these, our approach successfully
classifies all events as anomalous\footnote{Two events corresponded to direct
blocking of Tor bridge nodes, and these were identified as anomalous in the
bridge usage statistics. All other anomalies were detected in normal Tor
usage.}. In all cases except the Iranian DPI filtering on TLS that occurred in
2012, our anomalies coincide with the reported event from \cite{tor-timeline}.
In the case of Iran in 2012, we detect an anomalous period beginning two weeks
\textit{before} the reported event, corresponding to an immediate sharp fall in
Tor usage, followed by a longer period of slow decline over the following
month. 


\begin{table}
	\begin{center}
		\begin{threeparttable}
			\begin{tabulary}{\textwidth}{CLL}
				\hline
				Date &			Country &	Description of Event\\ 
				\hline
				2012-10-18 & 	Iran & 		TLS key exchange DPI\tnote{1}.     \\
				2012-12-16 & 	Syria & 		DPI on TLS renegotiation. 									\\
				2013-01-30 & 	Japan & 		Bridge blocked.												\\
				2013-03-09 & 	Iran & 		SSL handshake filtered.    				         	\\
				2013-03-26 & 	China & 		Probing obfs2 bridges.						     	\\
				2014-03-28 & 	Turkey & 	Tor website blocked.                           		\\
				2014-07-29 & 	Iran & 		Block directory authorities.            				\\
				2015-02-01 & 	China & 		Obfs4 bridges blocked.							\\
				\hline
			\end{tabulary}
			\begin{tablenotes}
			\item[1] See \S\ref{known-event-detection} for a discussion of this particular anomaly.
			\end{tablenotes}
		\end{threeparttable}
	\end{center}
	\caption{Complete list of reported, and detected, Tor blocking events.}
	\label{table:reported-events}
\end{table}

\subsection{Recent Events}


We have, in the course of investigating Tor metrics data with the tool detailed
in this work, discovered and reported a number of significant Tor usage
anomalies in countries including Ukraine, Israel, Bangladesh, UAE, and
Turkmenistan. In some of these cases anomalies are due to filtering behaviour,
such as Bangladesh's blocking of Facebook and chat applications in November
2015. In other cases the anomalies are due to external factors such as
Ukraine's blocking of the popular Russian social networking site VKontakte in
May 2017~\cite{guardian2017ukraine} that led to a large spike in circumvention
tool usage. Numerous other events have been detected, but space limitations
prevent significant discussion of individual cases.

\begin{figure}
	\centering
	\includegraphics[width=0.45\textwidth]{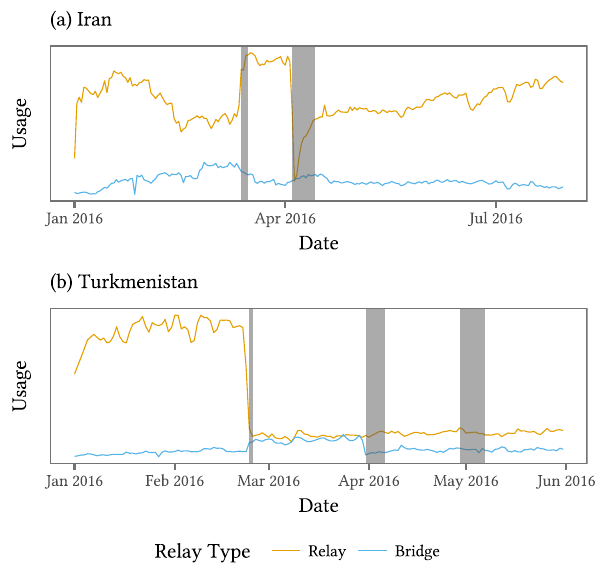}
	\caption{Combined relay and bridge Tor usage anomalies.}
	\label{figure:examples}
\end{figure}


\section{Example Results}
\label{section:examples}

Due to space constraints, we will not discuss specific cases in detail. This section shows a number of example outputs that highlight detected anomalies. As far as possible, we have extended the range of time shown in each plot to highlight that detected anomalies are not a frequent occurrence.

\subsection{Most Anomalous Countries}

Figure~\ref{fig:top-10} illustrates the ten most anomalous countries according
to their median absolute deviation from the median in the past year. Shaded
regions denote periods of anomalous usage, according to our tool.

%

\subsection{Combined Tor Metric Anomalies}

Figure~\ref{figure:examples} highlights example combined anomalies that demonstrate periods in which Tor usage via normal relays and access via bridge nodes experienced simultaneous but opposing anomalies.

Over the period included in the available Tor metrics data, which covers late 2011 until the time of writing, our technique identified 485 anomalous periods in which both Tor usage and bridge usage were jointly anomalous, across 102 countries out of the total 251 for which Tor assigns usage statistics. This number is somewhat inflated due to the fact that a number of these anomalous periods are separated only by a small number of days and are likely the result of the same event.

Of these countries, Georgia had the highest number of combined detected
anomalies, with 16 anomalous periods identified since 2011. The median number
of anomalous periods over the set of all 102 countries that showed any
anomalous behaviour was four. It is possible that this number may increase if
the combination of anomalous periods is made more flexible, as discussed in
\S\ref{section:combined}, however this demonstrates that events that exceed the
threshold for combined anomalies are relatively rare.

\subsection{Ukraine Russian Service Ban}

\begin{figure}
	\centering
	\includegraphics[width=0.45\textwidth]{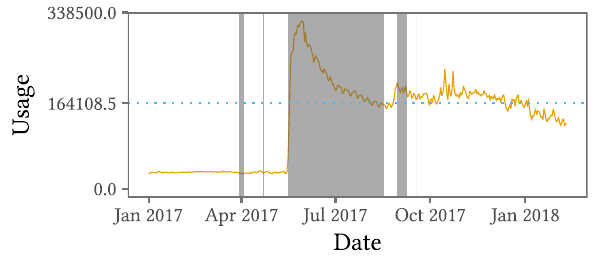}
	\caption{Anomalous usage following Ukraine's ban on major Russian network services.}
	\label{fig:ukraine}
\end{figure}

In early May 2017 the Ukrainian government blocked a number of major Russian online services, used by a significant number of Ukrainian citizens, including social network sites VKontakte and Odnoklassniki, mail provider mail.ru, and Yandex, a major search engine\cite{guardian2017ukraine}. Figure~\ref{fig:ukraine} shows a strong surge in Tor usage in the immediate aftermath of this, causing Ukraine to rise to the top of the daily anomaly rankings on the \texttt<name-redacted> mailing list. This example represents a significant anomaly in Tor usage related to blocking of standard internet services beyond Tor, and is in direct comparison to the Turkmenistan example of Figure~\ref{figure:examples} that highlights blocking of the Tor network itself.

%
%
%


\section{Discussion}

%
The validation and results of \S{\ref{section:validation}} and
\S{\ref{section:examples}} demonstrate that our approach is practically useful
for identifying both Tor blocking and, more generally, for identifying periods
of anomalous Tor usage. The highlighted anomalies detected by our approach
are strong indicators of regions of likely interest to the internet filtering
research and activist communities, and in particular in the combination of
normal Tor and bridge node usage.

More directly, the experimental validation in the previous section demonstrates
that our approach does detect a significant number of anomalies with varying
magnitudes and durations. 

\section{Future Work}
\label{section:future-work}


A main aspect of future work, for which these techniques were developed, will
be to perform analysis on historical filtering behaviour and to maintain an
ongoing watch for new potential filtering events. By combination with datasets
such as Google's Global Database of Events, Language, and Tone
(GDELT)~\cite{Leetaru13gdelt:global}, and through collaboration with
researchers and activists, the authors hope to develop and maintain a
contextualised time series of per-country filtering events for the benefit of
future researchers.

Whilst the work presented here has focused on the application of our technique
to Tor metrics data, the method is more generally applicable. Applying the
techniques presented here to other data sources is the most obvious direct
extension to this work. We have made preliminary analyses based on data from
Psiphon, CAIDA, Measurement Lab~\cite{mlab}, and the Wikimedia Foundation, as
well as evaluating data from the OONI Project~\cite{ooni} for its applicability
in detecting filtering. Other data sources, such as social media, are also
likely candidates for analysis. 


Given the results of combining multiple Tor metrics, an interesting line of
enquiry would be to investigate the speed with which users respond to filtering
of Tor by adopting bridge nodes, and to understand the proportion of users that
make this change. As more data sources are combined, further analysis of
filtering's effects in different countries and under different conditions
becomes possible.



\section{Conclusions}

We have developed a principal component analysis-based multivariate anomaly
detection system to detect anomalous periods in per-country usage statistics of
Tor metrics data. Our approach allows detection of per-country anomalies in
time series that are non-stationary and that demonstrate significant
seasonality. Our approach discounts global trends and even large-scale global
events by considering individual countries' usage patterns as relative to that
of others.

We have demonstrated the application of this tool to data from the Tor
Project's metrics portal, showing that it provides a means to indicate
potential censorship-related events, and others, at the global level. We have
further shown that combining multiple metrics to identify jointly-anomalous
periods can greatly improve the usefulness of the detected anomalies for
identifying periods of direct blocking of Tor.

This work presents a generally applicable tool for detecting a broad class of
internet filtering events on a global scale, without the need to focus on
individual countries, and that dynamically adapts to changing patterns of
usage. Countries exhibiting anomalous behaviour are automatically identified,
and can be subjected to further, more targeted, investigation.


We have validated our approach both by evaluating detection rates of injected
anomalies in a synthetically-generated time series, and demonstrated that our
detection rates are significantly higher than those used in the existing
anomaly detector used by the Tor project. Additionally, our tool provides
useful ranking of anomalies according to strength, as well as highlighting
anomalous periods rather than single-day events.

We have further evaluated our tool by successfully comparing detected anomalous
periods with an external list of known Tor blocking events. This evaluation
successfully identified each reported blocking event, supporting the tool's
practical effectiveness in detecting real-world anomalies.

Using our approach, we have demonstrated that combining anomalies detected in
multiple metrics can be an effective means to identify more targeted forms of
anomaly that indicate filtering behaviour. Our initial combination of
opposite-signed normal Tor usage and bridge node usage anomalies is a key step,
but there are other behaviours that could be of specific interest; there is
also significant potential for further combination with metrics from other
tools and data sources. 


Beyond the technique itself, the analyses presented in this work have
identified several states that are known to engage in active filtering, but
have also highlighted patterns of anomalous behaviour in several states that
have not received significant attention from the internet censorship research
community. Conducting more detailed investigations of these countries is a
promising focus for future research.

Our anomaly detection tool is running actively on a nightly basis, with results
output to a dedicated anomaly mailing list. This list has an audience amongst
NGOs and research projects working in the field of investigating filtering and
circumventing censorship, and has seen active use in detecting emerging
real-world filtering events.

In addition to the underlying technique and tool developed to detect anomalous
periods of behaviour, we have suggested, and provided initial evidence, that the
use of the Tor metrics data, amongst other sources, is of use not only as an
indicator of its own usage patterns, but as a practical proxy variable for a
much wider class of political and social events. This presents significant
potential for researchers, policy makers, and activists investigating global
freedom of expression.

\section{Acknowledgements}

This work was supported by \grantsponsor{ati}{The Alan Turing Institute}{https://www.turing.ac.uk} under the \grantsponsor{epsrc}{EPSRC}{https://www.epsrc.ac.uk} grant \grantnum{epsrc}{EP/N51012}. Joss Wright is partially funded by the Alan Turing Institute as a Turing Fellow under Turing Award Number \grantnum{ati}{TU/B/000044}.


\appendix

\bibliographystyle{ACM-Reference-Format}
\bibliography{censorship_detection_pca}

\end{document}